\documentclass[a4paper,aps,prd,twocolumn,preprintnumbers,showpacs,superscriptaddress,nofootinbib]{revtex4}
\usepackage{graphics,graphicx,epsfig}
\usepackage{amsfonts,amsmath,amssymb}

\input{tcilatex}

\begin{document}

\title{Boundary stress tensor and asymptotically AdS$_3$ non-Einstein \\
spaces at the chiral point}
\author{Gaston Giribet}
\email{gaston-at-df.uba.ar}
\affiliation{Universidad de Buenos Aires FCEN-UBA and IFIBA-CONICET, Ciudad
Universitaria, Pabell\'on I, 1428, Buenos Aires.}
\author{Andr\'es Goya}
\email{af.goya-at-df.uba.ar}
\affiliation{Universidad de Buenos Aires FCEN-UBA and IFIBA-CONICET, Ciudad
Universitaria, Pabell\'on I, 1428, Buenos Aires.}
\author{Mauricio Leston}
\email{mauricio-at-iafe.uba.ar}
\affiliation{Instituto de Astronom\'{\i}a y F\'{\i}sica del Espacio IAFE-CONICET, Ciudad
Universitaria, Pabell\'on IAFE, 1428 C.C. 67 Suc. 28, Buenos Aires.}

\begin{abstract}
Chiral gravity admits asymptotically AdS$_3$ solutions that are not locally
equivalent to AdS$_3$; meaning that solutions do exist which, while obeying
the strong boundary conditions usually imposed in General Relativity, happen
not to be Einstein spaces. In Topologically Massive Gravity (TMG), the
existence of non-Einstein solutions is particularly connected to the
question about the role played by complex saddle points in the Euclidean
path integral. Consequently, studying (the existence of) non-locally AdS$_3$
solutions to chiral gravity is relevant to understand the quantum theory.
Here, we discuss a special family of non-locally AdS$_3$ solutions to chiral
gravity. In particular, we show that such solutions persist when one deforms
the theory by adding the higher-curvature terms of the so-called New Massive
Gravity (NMG). Moreover, the addition of higher-curvature terms to the
gravity action introduces new non-locally AdS$_3$ solutions that have no
analogues in TMG. Both stationary and time-dependent, axially symmetric
solutions that asymptote AdS$_3$ space without being locally equivalent to
it appear. Defining the boundary stress-tensor for the full theory, we show
that these non-Einstein geometries have associated vanishing conserved
charges.
\end{abstract}

\maketitle

\section{Introduction}

In the last three years there has been an interesting discussion of whether
formulating three-dimensional Topologically Massive Gravity \cite{TMG} about
AdS$_{3}$ and Warped-AdS$_{3}$ spaces yields consistent models of quantum
gravity or not. The discussion started with the proposals in \cite{CG} and 
\cite{WAdS3}, and rapidly turned into a discussion on the consistency (the
closure) of imposing strong asymptotic boundary conditions that, by
extirpating undesired ghostly graviton excitations, would finally render the
background stable and the theory sensible at semiclassical level. In the
case of chiral gravity \cite{CG}, the specific discussion regarded the
question as to whether Brown-Henneaux boundary conditions \cite{BH}, usually
considered in Einstein gravity, are or not consistent in Topologically
Massive Gravity (TMG) at the chiral point ($\mu l=1$)\ as well. It was
argued in \cite{MSS} that TMG at $\mu l=1$ with the Brown-Henneaux boundary
conditions is actually a consistent model.

Apart from being essential to conclude the consistency (the stability) of
the theory (about a given background), the discussion on the boundary
conditions is also crucial to establish which are the geometries that
ultimately contribute to the partition function of the quantum theory. A
precise characterization of the set of geometries one has to consider has
not yet been accomplished. For example, questions like whether or not one
has to take non-smooth spaces into account are still unclear. Nevertheless,
with the aim of coming up with a physically sensible proposal for the
partition function, several assumptions and conjectures have been made about
over which geometrical configurations one has to sum. For instance, in the
case of chiral gravity, it has been conjectured that, after imposing strong
boundary conditions, all the saddle point contributions to the partition
function would come from Einstein spaces, and, consequently, would be
characterizable in terms of quotients of AdS$_{3}$ \cite{MaloneyWitten},
which would result in a remarkable simplification. This conjecture was
particularly expressed in \cite{MSS}, where it was mentioned that the fact
that at linearized level any solution of chiral gravity is locally
equivalent to AdS$_{3}$ might lead one to suspect that something similar
could happen at the full non-linear level. The suspicion was partially
supported by the observation that all stationary, axially symmetric
solutions of chiral gravity are indeed the solutions of General Relativity
(GR). However, it has been subsequently observed that less symmetric
non-Einstein spaces obeying the strong asymptotic conditions also exist in
Topologically Massive Gravity at $\mu l=1$. Then, solutions that are not
locally equivalent to AdS$_{3}$ (hereafter called non-locally AdS$_{3}$
solutions) have probably to be taken into account as well.

In chiral gravity, the question about the existence of non-locally AdS$_{3}$
solutions is particularly connected to the question about the role played by
complex saddle points in the Euclidean path integral. It was noted in \cite%
{MSS} that non-Einstein Lorentzian solutions of chiral gravity would yield
complex saddle points in the Euclidean theory. The argument goes as follows:
When rotating to Euclidean signature, the Cotton tensor $C_{\mu \nu }$ picks
up an imaginary factor $i$, so that the equations of motion of chiral
gravity take the form%
\begin{equation}
G_{\mu \nu }-\frac{1}{l^{2}}g_{\mu \nu }+ilC_{\mu \nu }=0,  \label{EoM}
\end{equation}%
with $G_{\mu \nu }=R_{\mu \nu }-\frac{1}{2}g_{\mu \nu }R$ being the Einstein
tensor. Then, since $C_{\mu \nu }$ vanishes for all Einstein spaces, all the
solutions to (\ref{EoM}) that are not solutions to GR would correspond to
complex saddle points in the Euclidean theory.

Then, the question arises as to how to deal with the contribution coming
from non-Einstein spaces in the theory. Do the complex saddle points
actually contribute to the Euclidean functional? A radical way to address
the problem (instead of trying to account for non-Einstein contributions)
would be trying to deform the theory in a way that, while still keeping the
desired properties of chiral gravity, non-locally AdS$_{3}$ spaces
ultimately result excluded. The first proposal to do this would be adding to
the Chern-Simons gravitational term of the TMG Lagrangian the special
combination of square-curvature terms proposed in the New Massive Gravity
(NMG) model of \cite{NMG}, which seems to be the natural candidate to yield
a consistent generalization of chiral gravity. However, as we will see, this
is not sufficient to exclude non-Einstein spaces from the asymptotically AdS$%
_{3}$ sector; such spaces actually persist. Moreover, the higher-order
corrections introduce additional solutions of such kind. All these
solutions, however, seem to exhibit some sort of pathologies; they present
closed timelike curves or timelike geodesic incompleteness in naked regions.
Nevertheless, since it is not clear whether summing only over smooth
geometries in the Euclidean path integral is the appropriate way of defining
the quantum theory, the existence of these non-locally AdS$_{3}$ spaces
deserves attention. Then, it seems reasonable to try to live in harmony with
the non-Einstein contributions, and explore the implications of taking these
geometries into account. Here, we will study a particular family of
non-Einstein spaces that asymptote AdS$_{3}$. In particular, we will show
that the solutions found in \cite{Detournay} can be extended to the model
consisting of coupling both TMG and NMG at the chiral point (interpolating
in such a way with solutions discussed in \cite{GiribetLeston}). We will
further generalize these solutions by finding asymptotically AdS$_{3}$
axially symmetric deformations of the extremal Ba\~{n}%
ados-Teitelboim-Zanelli black hole (BTZ). Defining the boundary
stress-tensor and resorting to standard holographic renormalization
techniques, we will compute the conserved charges of these asymptotically AdS%
$_{3}$ solutions and show they have associated vanishing mass and vanishing
angular momentum.

\section{Massive gravity in AdS$_{3}$}

\subsection{The action}

Let us begin by reviewing three-dimensional massive gravity about
asymptotically AdS$_{3}$ spaces. The action of the theory can be written as
the sum of four distinct contributions, namely%
\begin{equation}
S=S_{\text{EH}}+S_{\text{CS}}+S_{\text{NMG}}+S_{\text{B}},  \label{action}
\end{equation}%
where the first term corresponds to the Einstein-Hilbert action of GR 
\begin{equation}
S_{\text{EH}}=\frac{1}{16\pi G}\int_{\Sigma }d^{3}x\sqrt{-g}\left(
R-2\lambda \right) ,
\end{equation}%
the second term is the gravitational Chern-Simons term of TMG \cite{TMG} 
\begin{equation}
S_{\text{CS}}=\frac{1}{32\pi G\mu }\int_{\Sigma }d^{3}x\varepsilon ^{\alpha
\beta \gamma }\Gamma _{\alpha \sigma }^{\rho }(\partial _{\beta }\Gamma
_{\gamma \rho }^{\sigma }+\frac{2}{3}\Gamma _{\beta \eta }^{\sigma }\Gamma
_{\gamma \rho }^{\eta }),  \label{STMG}
\end{equation}%
and the third term contains the square-curvature contributions 
\begin{equation}
S_{\text{NMG}}=\frac{1}{16\pi Gm^{2}}\int_{\Sigma }d^{3}x\sqrt{-g}(R_{\mu
\nu }R^{\mu \nu }-\frac{3}{8}R^{2}),  \label{SNMG}
\end{equation}%
proposed in \cite{NMG}. The action also includes boundary terms $S_{\text{B}%
} $, which will be specified below.

The field equations derived from (\ref{action}) read 
\begin{equation}
G_{\mu \nu }+\lambda g_{\mu \nu }+\frac{1}{\mu }C_{\mu \nu }+\frac{1}{2m^{2}}%
K_{\mu \nu }=0.  \label{eom}
\end{equation}%
which, apart from the Einstein tensor $G_{\mu \nu }$ and the cosmological
constant term, include the Cotton tensor 
\begin{equation}
C_{\mu \nu }=\frac{1}{2}\varepsilon _{\mu }^{\ \text{\ }\alpha \beta }\nabla
_{\alpha }R_{\beta \nu }+\frac{1}{2}\varepsilon _{\nu }^{\ \text{\ }\alpha
\beta }\nabla _{\alpha }R_{\mu \beta },  \label{Cotono}
\end{equation}%
and the tensor $K_{\mu \nu }$ 
\begin{eqnarray}
K_{\mu \nu } &=&2\square {R}_{\mu \nu }-\frac{1}{2}\nabla _{\mu }\nabla
_{\nu }{R}-\frac{1}{2}\square {R}g_{\mu \nu }+4R_{\mu \alpha \nu \beta
}R^{\alpha \beta }  \notag \\
&&-\frac{3}{2}RR_{\mu \nu }-R_{\alpha \beta }R^{\alpha \beta }g_{\mu \nu }+%
\frac{3}{8}R^{2}g_{\mu \nu }.  \label{Kaono}
\end{eqnarray}%
The latter satisfies the remarkable property%
\begin{equation}
g^{\mu \nu }K_{\mu \nu }=R_{\mu \nu }R^{\mu \nu }-\frac{3}{8}R^{2},
\end{equation}%
which is one of the reasons why the theory proposed in \cite{NMG} is a very
special one. The theory has three mass scales, namely $\mu ,$ $m$, $|\lambda
|^{1/2}$. While the case $1/\mu =0$ yields NMG, the case $1/m^{2}=0$
corresponds to TMG. Here, we will be concerned with the full theory (\ref%
{action}).

Three-dimensional massive gravity (\ref{action}) admits AdS$_{3}$ space as a
solution. In fact, it is not hard to verify that all Einstein spaces satisfy
(\ref{eom}). The two admissible values for the typical radius of the AdS$%
_{3} $ solution are given by%
\begin{equation}
l^{2}=-\frac{1}{2\lambda }\left( 1\pm \sqrt{1+\lambda m^{-2}}\right) ,
\label{Radius}
\end{equation}%
as it can be easily seen from the trace of the equations of motion (\ref{eom}%
). If $\lambda <0$ and $1+\lambda m^{2}>0$, AdS$_{3}$ space exists as
solution.

\subsection{Boundary conditions}

Since we are originally motivated by AdS$_{3}$/CFT$_{2}$, we will be
concerned with spaces that asymptote AdS$_{3}$ near its boundary. Then, the
first question that appears is which definition of 'asymptotically AdS$_{3}$
spaces' we have to take into account in the theory (\ref{action}). The
asymptotic-AdS$_{3}$ conditions are defined by requiring the conserved
charges computed in the boundary of the space to be finite, but in a way
that is still compatible with a sufficiently interesting subset of the space
of solutions. The question about the asymptotic is important because, as it
has been observed in different scenarios \cite{boundaryconditions1,
boundaryconditions2}, the adequate definition of asymptotic-AdS$_{3}$
conditions is a theory-dependent notion. Besides, the consistency of a given
set of boundary conditions not only depends on the specific Lagrangian, but
it also depends on the particular point of the parameter space in which one
is interested. For instance, both TMG\ and NMG\ exhibit special points in
the parameter space at which AdS$_{3}$ asymptotic boundary conditions can be
defined with a falling-off behavior that results relaxed with respect to
that in Einstein gravity. It was shown in \cite{Sun} that for generic values
of $m^{2}l^{2}$ the appropriate asymptotic conditions to be considered in
NMG\ are the Brown-Henneaux boundary conditions of \cite{BH}, the same as in
GR. In the system of coordinates in which the metric of (the universal
covering of) AdS$_{3}$ space takes the form%
\begin{equation}
ds^{2}=-\left( \frac{r^{2}}{l^{2}}+1\right) dt^{2}+\left( \frac{r^{2}}{l^{2}}%
+1\right) ^{-1}dr^{2}+r^{2}d\varphi ^{2},  \label{AdS3}
\end{equation}%
with $\varphi \in \lbrack 0,2\pi ]$, $t\in \mathbb{R}$, $r\in \mathbb{R}%
_{\geq 0}$, the Brown-Henneaux boundary conditions read 
\begin{eqnarray}
g_{tt} &=&-\frac{r^{2}}{l^{2}}+\mathcal{O}(1),\qquad g_{rt}=\mathcal{O}%
(r^{-4}),  \label{q1} \\
g_{rr} &=&\frac{l^{2}}{r^{2}}+\mathcal{O}(r^{-4}),\qquad g_{\varphi t}=%
\mathcal{O}(1), \\
\qquad g_{\varphi \varphi } &=&r^{2}+\mathcal{O}(1),\qquad g_{r\varphi }=%
\mathcal{O}(r^{-4}),  \label{q7}
\end{eqnarray}%
where $\mathcal{O}(r^{-n})$ stands for terms whose $r$-dependence damps off
as $1/r^{n}$ or faster at large $r$, with their dependence on the
coordinates $t$ and $\varphi $ being arbitrary. Boundary conditions (\ref{q1}%
)-(\ref{q7}) are consistent for the full theory (\ref{action}), in the sense
it yields finite charges.

The group of asymptotic Killing vectors preserving (\ref{q1})-(\ref{q7}) is
generated by two copies of the DeWitt algebra \cite{BH}, which certainly
includes the isometry algebra of AdS$_{3}$ as a proper subalgebra. More
interesting is the fact that, as it happens in three-dimensional Einstein
gravity \cite{BH}, the algebra satisfied by the conserved charges associated
to those asymptotic symmetries turns out to coincide with two copies of the
Virasoro algebra with left and right central charges%
\begin{eqnarray}
c_{L} &=&\frac{3l}{2G}\left( 1-\frac{1}{\mu l}+\frac{1}{2m^{2}l^{2}}\right) ,
\label{clcr} \\
c_{R} &=&\frac{3l}{2G}\left( 1+\frac{1}{\mu l}+\frac{1}{2m^{2}l^{2}}\right) ,
\end{eqnarray}%
respectively. From the AdS$_{3}$/CFT$_{2}$ perspective, these central
charges acquire the interpretation as being the central charges of the dual
conformal field theory (CFT). We will focus our attention on chiral gravity,
namely on the theory defining on the line $c_{L}=0$ of the parameter space
and by imposing Brown-Henneaux boundary conditions on the space of solutions.

In contrast with the chiral gravity of \cite{CG}, where $\mu l$ gets fixed
to $1$ by the requirement $c_{L}=0$, the theory defined by action (\ref%
{action}) at $c_{L}=0$ has the coupling of the higher-curvature terms as a
free parameter to play with. This will introduces more diversity in the kind
of asymptotically AdS$_{3}$ solutions we are interested in.

\section{Non-locally AdS$_{3}$ solutions}

\subsection{Persistent solutions}

Three-dimensional massive gravity (\ref{action}) admits all GR solutions as
exact solutions. This is evident in the case of TMG, as the Cotton tensor
vanishes if and only if a space is conformally flat. If the NMG contribution
is also included in the action this is still true. However, the reciprocal
is not true; that is, there also exist a rich set of solutions to (\ref%
{action}) that are not solutions to GR. Finding such a non-Einstein solution
is not necessarily a hard problem; what is a hard problem is to try to
answer the question as to whether or not non-Einstein solutions persist
after strong boundary conditions like (\ref{q1})-(\ref{q7}) are imposed. It
was shown recently that such solutions actually exist both in the case of
TMG\ \cite{Detournay} and in the case of NMG \cite{GiribetLeston}, and here
we will show that these solutions also exist in the general theory (\ref%
{action}) when $c_{L}=0$. To see this, let us start by considering the 
\textit{ansatz}%
\begin{equation*}
ds^{2}=\frac{l^{2}}{r^{2}}dr^{2}+\frac{r^{2}}{l^{2}}(-dt^{2}+l^{2}d\varphi
^{2})+\sum\nolimits_{\pm }h_{\pm \pm }(dt\pm ld\varphi )^{2},
\end{equation*}%
and consider the expansion%
\begin{equation}
h_{\pm \pm }(t,\varphi ,r)=h_{\pm }^{(0)}+r^{-2}h_{\pm }^{(2)}+r^{-4}h_{\pm
}^{(4)}+r^{-6}h_{\pm }^{(6)}\ ...  \label{Exp}
\end{equation}%
with $h_{\pm }^{(2k)}$ being functions of coordinates $t$ and $\varphi $.
The solution found in \cite{Detournay} corresponds to $h_{--}(t,\varphi
,r)=0 $, with $h_{+}^{(0)}\sim t$ and $h_{+}^{(4)}\sim const$. As in \cite%
{Detournay, GiribetLeston}, one may considers the \textit{ansatz}%
\begin{equation}
h_{--}(t,\varphi ,r)=0,\quad \quad h_{++}(t,r)=\frac{\beta }{l^{2}}t-\frac{%
\beta ^{2}l^{2}}{96n}r^{-4}  \label{ansatz}
\end{equation}%
and look for exact solutions. It turns out that, without major
modifications, equations (\ref{eom}) are solved for (\ref{ansatz})\ if and
only if 
\begin{equation*}
\mu =\dfrac{2m^{2}l}{2m^{2}l^{2}+1},\qquad \lambda =\dfrac{1-4m^{2}l^{2}}{%
4m^{2}l^{4}},\qquad n=\dfrac{2m^{2}l^{2}-11}{2m^{2}l^{2}-15},
\end{equation*}%
or, equivalently, if 
\begin{equation*}
2m^{2}l^{2}=\frac{\mu l}{1-\mu l},\qquad \lambda =\dfrac{1-3\mu l}{2\mu l}%
,\qquad n=\dfrac{12\mu l-11}{16\mu l-15}.
\end{equation*}%
$\beta $ is an arbitrary constant. It is easy to check that these values of
the parameters precisely correspond to the following values of the central
charges%
\begin{equation}
c_{L}=0,\qquad \qquad c_{R}=\frac{3l}{G},
\end{equation}%
which is a generalization of the chiral point studied in \cite{CG}. That is,
the generalized chiral gravity, defined as considering $2m^{2}l^{2}=\mu
l/(1-\mu l)$ in (\ref{action}) and imposing the asymptotic behavior (\ref{q1}%
)-(\ref{q7}), exhibits solutions that are not solutions of general
relativity. We will find more general solutions of this type below.

Space (\ref{ansatz}) is a time-dependent geometry that, in spite of it,
still admits $\partial _{t}$ as asymptotic timelike Killing vector, in the
sense that it satisfies boundary conditions (\ref{q1})-(\ref{q7}). This
provides us with a notion of gravitational mass and angular momentum for
this space. Being a genuine asymptotically AdS$_{3}$ configuration, (\ref%
{ansatz}) deserves to be studied and it is worthwhile asking ourselves about
its implications for AdS$_{3}$/CFT$_{2}$. Geometrical aspects of space (\ref%
{ansatz}) have been analyzed in \cite{Detournay}, where it was observed that
it exhibits closed timelike curves at the time dependent radius $r_{\text{ctc%
}}$ that solves the equation $r_{\text{ctc}}+l^{2}h_{++}(t,r_{\text{ctc}})=0$%
. Curvature invariants of this space are%
\begin{equation*}
R=-\frac{6}{l^{2}},\qquad R_{\mu \nu }R^{\mu \nu }=\frac{12}{l^{4}},\qquad
R_{\mu \alpha \nu \beta }R^{\mu \nu }R^{\alpha \beta }=-\frac{24}{l^{6}}.
\end{equation*}%
The fact that these curvature invariants coincide with those of AdS$_{3}$
space may suggest that (\ref{ansatz}) correspond to a locally AdS$_{3}$
space; however, this is not the case. In fact, the Cotton tensor $C_{\mu \nu
}$ associated to (\ref{ansatz}) does not vanish unless $\beta =0$, which
implies that the space is not conformally flat, and thus it is not locally
equivalent to AdS$_{3}$. Namely,%
\begin{equation*}
C_{\mu \nu }=\frac{\beta }{r^{4}}\left( 
\begin{array}{ccc}
\frac{1}{8}\beta (4-5/n)/l & \frac{1}{8}\beta (4-5/n) & r/l \\ 
\frac{1}{8}\beta (4-5/n) & \frac{1}{8}\beta l(4-5/n) & r \\ 
r/l & r & 0%
\end{array}%
\right) ,
\end{equation*}%
with the labels $x^{0}=t$, $x^{1}=\varphi $, $x^{2}=r$.

One can in principle solve the equations of motion iteratively considering
expansion (\ref{Exp}). Solution (\ref{ansatz}) corresponds to the case $%
h_{--}(t,\varphi ,r)=0$ and $h_{++}^{(0)}(t,\varphi ,r)=\beta
(t-t_{0})/l^{2}-\beta ^{2}l^{2}/(96nr^{4})$. It still remains a solution if
a quadratic piece $\sim r^{2}$ is added to this function; however, in that
case conditions (\ref{q1})-(\ref{q7}) are not obeyed.

\subsection{New solutions}

One might ask whether one can generalize (\ref{ansatz}) in a way that
Brown-Henneaux asymptotic is preserved. To see that this is actually
possible one has to consider a more general solution whose expansion (\ref%
{Exp}) is not necessarily finite. Such a solution is given by%
\begin{equation}
h_{++}(t,r)=h_{++}^{(0)}(t,r)-h_{+}^{(L)}\log
(r)+h_{+}^{(m)}r^{3/2-m^{2}l^{2}},  \label{generalized}
\end{equation}%
where $h_{+}^{(L)}$ and $h_{+}^{(m)}$ are two arbitrary constant
coefficients. This generalizes (\ref{ansatz}) and obeys the Brown-Henneaux
conditions if $h_{+}^{(L)}=0$ and $m^{2}l^{2}\geq 3/2$. The solution with $%
h_{+}^{(L)}\neq 0$, on the other hand, reduces to the Log-gravity solution
studied in \cite{GAY} when $\beta =h_{+}^{(m)}=0$, and even if it does not
obey Brown-Henneaux boundary conditions (\ref{q1})-(\ref{q7}), it happens to
be asymptotically AdS$_{3}$ in the sense of \cite{boundaryconditions2}.

Solution (\ref{generalized}) is not locally AdS$_{3}$ for generic values of
the coefficients. In fact, for (\ref{generalized}) to be conformally flat it
is necessary (not sufficient) to have $\beta =0$. Besides, even it the $t$%
-dependent term is not present, the Cotton tensor takes the form

\begin{equation*}
C_{\mu \nu }=\left( h_{+}^{(L)}+\frac{P(m^{2}l^{2})}{16r^{m^{2}l^{2}-3/2}}%
h_{+}^{(m)}\right) \left( 
\begin{array}{ccc}
1/l^{3} & 1/l^{2} & 0 \\ 
1/l^{2} & 1/l & 0 \\ 
0 & 0 & 0%
\end{array}%
\right) .
\end{equation*}%
with $P(x)=3-2x-12x^{2}+8x^{3}$. This gives three roots for the Cotton
tensor to vanish when $h_{+}^{(L)}=0$; namely, it only vanishes for $%
m^{2}l^{2}=\pm 1/2$ and $m^{2}l^{2}=3/2$ (and, of course, for $%
m^{2}l^{2}=\infty $). The points $m^{2}l^{2}=\pm 1/2$ always exhibit special
features in NMG. Another special point is $m^{2}l^{2}=15/2$, where $n$
diverges. There, only a damping $\sim 1/r^{6}$ is present, namely one finds $%
h_{+}^{(6)}\neq 0$ with $h_{+}^{(4)}=0$ in (\ref{Exp}). Notice also that the
term $\sim 1/r^{m^{2}l^{2}-3/2}$ in (\ref{generalized}) coincides with the $%
\sim 1/r^{4}$ dependence of $h_{++}^{(0)}(t,r)$ for $m^{2}l^{2}=11/2$, where 
$n$ vanishes. There, one can set $\beta ^{2}\propto n$ so that the
time-dependent term disappears by keeping the term $\sim 1/r^{4}$ in the
metric. This represents a stationary, axially symmetric solution that
asymptotes AdS$_{3}$. More generically, the existence of solution (\ref%
{generalized}) already shows that Birkhoff-like theorems that were proven
for TMG do not hold when the higher terms of NMG are added to the action. In
fact, for $\beta =0$ solution (\ref{generalized}) is an asymptotically AdS$%
_{3}$ space which is stationary, and even when it has the same scalar
curvature that AdS$_{3}$ space, it is not locally equivalent to it.

\subsection{Perturbing the extremal black hole}

It is relatively easy to generalize (\ref{generalized}) further by showing
that a similar perturbation of the extremal BTZ is admitted as exact
solution to (\ref{eom}) when $c_{L}=0$. Consider the extremal BTZ metric%
\begin{equation}
ds^{2}=-N(r)dt^{2}+\frac{dr^{2}}{N(r)}+r^{2}\left( d\varphi +N^{\varphi
}(r)dt\right) ^{2},  \label{BTZ}
\end{equation}%
with 
\begin{equation}
N(r)=\frac{r^{2}}{l^{2}}-4M+\frac{4M^{2}l^{2}}{r^{2}},\qquad N^{\varphi }(r)=%
\frac{2Ml}{r^{2}}.  \label{BTZ2}
\end{equation}%
This geometry represents an extremal black hole \cite{BTZ}, which locally
equivalent to AdS$_{3}$. The event horizon of the black hole is located at $%
r_{+}=l\sqrt{2M}$. It turns out that, by adding to metric (\ref{BTZ})-(\ref%
{BTZ2}) a term $h_{++}(dt+ld\phi )^{2}$ with 
\begin{equation}
h_{++}(r)=-\frac{1}{2}h_{+}^{(L)}\log
(r^{2}-r_{+}^{2})+h_{+}^{(m)}(r^{2}-r_{+}^{2})^{\frac{3-2m^{2}l^{2}}{4}},
\label{hBH}
\end{equation}%
gives a new solution to (\ref{eom}) when $c_{L}=0$ (again, $h_{+}^{(L)}$ and 
$h_{+}^{(m)}$ are two arbitrary constants.) This new solution generalizes
the logarithmic deformation of the BTZ solution found in \cite{GAY}. This
obeys the weakened asymptotic%
\begin{eqnarray*}
g_{tt} &=&-\frac{r^{2}}{l^{2}}+\mathcal{O}(\log (r)),\qquad g_{rt}=\mathcal{O%
}(r^{-4}), \\
g_{rr} &=&\frac{l^{2}}{r^{2}}+\mathcal{O}(r^{-4}),\qquad g_{\varphi t}=%
\mathcal{O}(\log (r)), \\
\qquad g_{\varphi \varphi } &=&r^{2}+\mathcal{O}(\log (r)),\qquad
g_{r\varphi }=\mathcal{O}(r^{-4}),
\end{eqnarray*}%
which are consistent with the boundary conditions in the definition of
Log-gravity \cite{boundaryconditions2, MSS}. On the other hand, when $%
h_{+}^{(L)}=0$ the logarithmic term in (\ref{hBH}) is not present, and the
solution happens to obey Brown-Henneaux boundary conditions (\ref{q1})-(\ref%
{q7}) with $h_{+}^{(m)}\neq 0$ if $m^{2}l^{2}\geq 3/2$. Metric function (\ref%
{hBH}) diverges at the 'would-be-horizon' radius $r_{+}=l\sqrt{2M}$, and
this divergence makes the geodesics of falling particles to wind infinitely
rapid before reaching this radius. It is also important to mention that for $%
r^{2}<2Ml$ the metric of the space can be extended by reversing the sign in
the argument of the logarithm in (\ref{hBH}).

The space described by the perturbation (\ref{hBH}) is not conformally flat.
In general, for a perturbation of the extremal BTZ\ metric of the form 
\begin{equation}
h_{++}(r)=-h_{+}^{(0)}\log
(r)+h_{+}^{(2)}r^{-2}+h_{+}^{(4)}r^{-4}+h_{+}^{(6)}r^{-6}\ ...
\end{equation}%
with the coefficients $h_{+}^{(2k)}$ now being constant, the non-vanishing
components of the Cotton tensor are proportional to%
\begin{equation}
\dfrac{r^{2}-r_{+}^{2}}{r^{6}}%
\sum_{k=0}Q_{2k}(r_{+}^{2},r^{2})r^{-2k}h_{+}^{(2k)},
\end{equation}%
with $Q_{2k}(r_{+}^{2},r^{2})$ being polynomials of the form 
\begin{eqnarray*}
Q_{0}(x,y) &=&(1y^{2}-5xy+4x^{2}), \\
Q_{2}(x,y) &=&4(3y^{2}-9xy+6x^{2}), \\
Q_{4}(x,y) &=&12(5y^{2}-13xy+8x^{2}), \\
Q_{6}(x,y) &=&24(7y^{2}-17xy+10x^{2}), \\
&&...
\end{eqnarray*}%
More precisely, one finds $%
Q_{2k}(x,y)=C_{2k}((2k+1)y^{2}-(4k+5)xy+(2k+4)x^{2})$, where $C_{0}=1,$ $%
C_{2}=4$, and $C_{2k}=C_{2k-2}+4k$ for $k>1$. In the large $r$ limit only
the logarithmic term gives a non-conformally flat contribution.

\subsection{The Poincar\'{e} patch}

Solution (\ref{ansatz}) can be generalized in a way that the metric acquires
linear dependence both in $t$ and $\varphi $. For the metric not to be
multivalued, one has to take the universal covering of coordinate $\varphi $%
; this is done by defining $x=l\varphi \in \mathbb{R}$. Defining light-cone
coordinates $x^{\pm }=t\pm x$ and the new coordinate $z=l^{2}/r$, we have
the more general solution%
\begin{equation}
ds^{2}=l^{2}\frac{dz^{2}-dx^{-}dx^{+}}{z^{2}}+\kappa h_{++}(dx^{+})^{2}
\label{anSa}
\end{equation}%
with%
\begin{equation}
h_{++}(x^{+},x^{-},z)=x^{+}+\sigma x^{-}-\frac{\kappa \sigma ^{2}}{24nl^{4}}%
z^{4}+c_{m}\ z^{m^{2}l^{2}-\frac{3}{2}},  \label{ansa2}
\end{equation}%
where we defined $\kappa =\beta /(2l^{2})$, we rescaled $h_{+}^{(m)}$ to
define $c_{m}$, and introduced a new parameter $\sigma $ which can be taken
to be $-1\leq \sigma \leq 1$. The case $\sigma =1$ is that of (\ref%
{generalized}). Space (\ref{anSa})-(\ref{ansa2}) is a perturbation of AdS$%
_{3}$ space written in Poincar\'{e} coordinates; the latter corresponds to $%
\kappa =0$. The spaces represented by metrics with different values of $%
\sigma $ are locally isometric.

To analyze the large $z$ limit, namely the region $r\approx 0$ of the space,
one may define a new variable $\zeta =z^{-2}$ and rescaling $x^{+}$ by a
constant factor, for $c_{m}=0$ we find that the leading piece of the metric
takes the form%
\begin{equation*}
ds^{2}\simeq \frac{l^{2}}{4\zeta ^{2}}\left( d\zeta ^{2}-(dx^{+})^{2}\right)
+\mathcal{O}(\zeta )dx^{+}dx^{-}+\mathcal{O}(1)(dx^{+})^{2}.
\end{equation*}%
This is the limit ($r\sim \zeta ^{1/2}\sim 0$) the timelike coordinate $t$
and the spacelike coordinate $r$ tends to a lightlike coordinate. This limit
was considered in \cite{Detournay} to discuss the nature of geodesic
incompleteness at $r\simeq 0$.

\section{Boundary stress tensor}

\subsection{The Brown-York tensor}

Now, let us discuss the definition of a boundary stress tensor and
holographic renormalization \cite{Skenderis, BK} for these spaces. The
general idea is that Brown-York tensor \cite{BY} applied to asymptotically
AdS spaces, and upon an appropriate regularization of the charges it yields
when evaluating at the boundary $r\rightarrow \infty $, gives the
stress-tensor of the dual conformal field theory (CFT). Schematically, one
can identify such renormalized version of the Brown-York tensor $T_{ij}^{(%
\text{ren})}$ with the quantity%
\begin{equation}
\lim_{r\rightarrow \infty }T_{ij}^{(\text{ren})}\equiv \left\langle
T_{ij}\right\rangle _{\text{CFT}}=\frac{2}{\sqrt{-\gamma }}\frac{\delta S_{%
\text{eff}}}{\delta \gamma ^{ij}}|_{\gamma _{ij}=\delta _{ij}}  \label{Tef}
\end{equation}%
where $S_{\text{eff}}$ refers to the effective action of the gravity theory
in AdS$_{3}$.

Adding boundary terms to the gravity action is necessary for the Brown-York
tensor to be finite when evaluated at the boundary $r=\infty $. Such
boundary terms are of two kinds:\ First, those terms $S_{\text{B}}$ needed
for the variational principle to be defined in an specific way; secondly,
those terms $S_{\text{C}}$ that, being constituted of intrinsic quantities
of the boundary, may be added to the action without changing the classical
dynamics and yielding finite conserved charges at the boundary. From the
dual theory point of view, the boundary terms $S_{\text{C}}$ in the action
can be thought of as local counterterms needed by renormalization.

To define the boundary stress-tensor it is convenient to write the metric in
its Arnowitt-Deser-Misner (ADM) decomposition 
\begin{equation}
ds^{2}=N^{2}dr^{2}+\gamma _{ij}(dx^{i}+N^{i}dr)(dx^{j}+N^{j}dr),  \label{ADM}
\end{equation}%
where $N^{2}$ is the radial lapse function, and $\gamma _{ij}$ is the
two-dimensional metric on the constant-$r$ surfaces. The Latin indices refer
to the coordinates on the constant-$r$ surfaces $i,j=0,1$, while the Greek
indices are $\mu ,\nu =0,1,2$, recall $x^{2}=r$.

\subsection{Boundary action}

Being a theory that yields fourth-order equations of motion, NMG makes the
discussion on boundary action a little bit more subtle; one has to give a
prescription for how the variational principle is to be defined, for how the
boundary data is to be specified. Here, we will follow the proposal in \cite%
{HohmTonni}, which amounts to first rewriting the action of NMG in an
alternative way by introducing an auxiliary field $f^{\mu \nu }$ and
defining the alternative action 
\begin{eqnarray}
S_{\text{A}} &=&\frac{1}{16\pi G}\int_{\Sigma }d^{3}x\sqrt{-g}\left(
R-2\lambda +f^{\mu \nu }(R_{\mu \nu }-\frac{1}{2}g_{\mu \nu }R)\right. 
\notag \\
&&\left. -\frac{1}{4}m^{2}(f_{\mu \nu }f^{\mu \nu }-f^{2})\right) +S_{\text{%
CS}}.  \label{SA}
\end{eqnarray}%
with $f=f_{\mu \nu }g^{\mu \nu }$. In fact, on-shell, $f_{\mu \nu }$ results
proportional to the Schouten tensor, namely 
\begin{equation}
f_{\mu \nu }=\frac{2}{m^{2}}(R_{\mu \nu }-\frac{1}{4}g_{\mu \nu }R),
\label{schotten}
\end{equation}%
and plug it back in (\ref{SA}) one recovers the bulk action in its original
form.

Then, boundary terms may be introduced for the variational principle to be
defined in such a way that both the metric $g_{\mu \nu }$ and the auxiliary
field $f_{\mu \nu }$ are fixed on the boundary. With this prescription, the
boundary action reads%
\begin{equation}
S_{\text{B}}=\frac{1}{16\pi G}\int_{\partial \Sigma }d^{2}x\sqrt{-\gamma }%
\left( -2K-\hat{f}^{ij}K_{ij}+\hat{f}K\right) ,  \label{Sb}
\end{equation}%
where $\gamma _{ij}$ is the metric induced on the boundary, $K_{ij}$ is the
extrinsic curvature, and $K$ is its trace $K=\gamma ^{ij}K_{ij}$. In ADM
variables, the extrinsic curvature takes the form $K_{ij}=-\frac{1}{2N}%
\left( \partial _{r}\gamma _{ij}-\nabla _{i}N_{j}-\nabla _{j}N_{i}\right) $.
For convenience, we can decomposed the auxiliary field $f^{\mu \nu }$ as
follows%
\begin{equation}
f^{\mu \nu }=\left( 
\begin{array}{cc}
f^{ij} & h^{i} \\ 
h^{j} & s%
\end{array}%
\right)
\end{equation}%
and define the combinations $\hat{f}^{ij}=f^{ij}+2h^{(i}N^{j)}+sN^{i}N^{j},$ 
$\hat{h^{i}}=N(h^{i}+sN^{i}),$ $\hat{s}=N^{2}s,$ $\hat{f}=\gamma ^{ij}\hat{f}%
_{ij}$.

The first term in (\ref{Sb}) is of course the Gibbons-Hawking term,
corresponding to the Einstein-Hilbert bulk action. The other two terms in (%
\ref{Sb}) are proper of NMG. No additional boundary terms are needed because
the Chern-Simons gravitational term of TMG is included in the action. Then,
the (unrenormalized) boundary stress tensor of the full theory is given by%
\begin{equation}
T_{ij}=\frac{2}{\sqrt{-\gamma }}\frac{\delta }{\delta \gamma ^{ij}}(S_{\text{%
A}}+S_{\text{B}}),  \label{Tij}
\end{equation}%
which, again, can be written identifying three distinct contributions,
namely $T^{ij}=T_{\text{EH}}^{ij}+T_{\text{CS}}^{ij}+T_{\text{NMG}}^{ij}$.
The first contribution would be the standard contribution coming from GR,

\begin{equation}
T_{\text{EH}}^{ij}=\frac{1}{8\pi G}(K^{ij}-K\gamma ^{ij}),  \label{TA}
\end{equation}%
the second one is the contribution corresponding to the Chern-Simons
gravitational term%
\begin{equation}
T_{\text{CS}}^{ij}=\frac{1}{16\pi G\mu }\left( \epsilon ^{k(i}\gamma
^{j)l}\partial _{r}K_{kl}+2\epsilon ^{k(i}\partial _{r}K_{k}^{j)}\right) ,
\label{TB}
\end{equation}%
where the gauge $N^{i}=0$, $N=1$ was chosen, such that $K_{ij}=-\frac{1}{2}%
\partial _{r}\gamma _{ij}$; and in the third place we have the contribution
coming from NMG, namely

\begin{equation*}
T_{\text{NMG}}^{ij}=-\frac{1}{8\pi G}\left( \frac{1}{2}\hat{f}K^{ij}+\nabla
^{(i}\hat{h}^{j)}-\frac{1}{2}\mathcal{D}_{r}\hat{f}^{ij}+K_{k}^{(i}\hat{f}%
^{j)k}\right.
\end{equation*}%
\begin{equation}
\quad \quad \left. -\frac{1}{2}\hat{s}K^{ij}-\gamma ^{ij}(\nabla _{k}\hat{h}%
^{k}-\frac{1}{2}\hat{s}K+\frac{1}{2}\hat{f}K-\frac{1}{2}\mathcal{D}_{r}\hat{f%
})\right) ,  \label{TC}
\end{equation}%
where the covariant $r$-derivatives $\mathcal{D}_{r}$, defined in \cite%
{HohmTonni}, in the gauge $N^{i}=0$, $N=1$ acts simply as an ordinary
derivative, namely $\mathcal{D}_{r}\hat{f}_{ij}=\partial _{r}\hat{f}_{ij}$, $%
\mathcal{D}_{r}\hat{f}=\partial _{r}\hat{f}$; see \cite{HohmTonni} for
details.

\subsection{Counterterms}

The next step to define the boundary stress-tensor is proposing the
counterterms in the action; namely, the additional boundary terms that
ultimately yield a regularized quantities in the boundary. This additional
boundary action $S_{\text{C}}$ has to be made of intrinsic boundary scalars
like%
\begin{equation}
S_{\text{C}}=\int d^{2}x\sqrt{-\gamma }\left( \alpha _{0}+\alpha _{1}\ \hat{f%
}+\alpha _{2}\ \hat{f}^{2}+\beta _{2}\ \hat{f}_{ij}\hat{f}^{ij}+...\right) ,
\label{SC}
\end{equation}%
where the ellipses stand for other local contributions.

As said before, from the boundary point of view, these terms are thought of
as counterterms in the dual CFT$_{2}$; meaning that the renormalized
boundary stress tensor is%
\begin{equation}
T_{ij}^{\text{(ren)}}=T_{ij}+\frac{2}{\sqrt{-\gamma }}\frac{\delta }{\delta
\gamma ^{ij}}S_{\text{C}}.  \label{Tijren}
\end{equation}

It is possible to verify that, for asymptotically AdS$_{3}$ spaces
satisfying Brown-Henneaux boundary conditions, it is sufficient to add the
term 
\begin{equation}
\alpha _{0}=-\frac{1}{8\pi Gl}\left( 1+\frac{1}{2m^{2}l^{2}}\right)
\label{tornadito}
\end{equation}%
for the stress-tensor to be finite in the large $r$ limit. That is, unlike
what happens in the case of weakened AdS$_{3}$ asymptotic \cite%
{GiribetLeston, delorto, Troncoso2} or in the case of non-AdS$_{3}$
asymptotic \cite{HohmTonni}, a boundary cosmological constant term is
sufficient to regularize the boundary tensor when Brown-Henneaux asymptotic
conditions are imposed. To see this, one may express the metric using
Fefferman-Graham expansion%
\begin{equation}
\gamma _{ij}=r^{2}\gamma _{ij}^{(0)}+\gamma _{ij}^{(2)}+r^{-2}\gamma
_{ij}^{(4)}+r^{-4}\gamma _{ij}^{(6)}...,  \label{FG}
\end{equation}%
where $\gamma _{ij}^{(2k)}$ are functions that only depend on $t$ and $%
\varphi $, and then explicitly verify that the value (\ref{tornadito}) in (%
\ref{SC}) with $\alpha _{i>0}=\beta _{i}=0$ yields finite components (\ref%
{Tijren}).

As a non-trivial check of the boundary stress-tensor proposed, let us
perform a simple calculation, let us compute the central charge $c=\frac{%
c_{L}+c_{R}}{2}$ through the trace anomaly calculation. The trace anomaly
and the diffeomorphisms anomaly calculation in presence of the gravitational
Chern-Simons term has been already discussed in the literature \cite{BK, KL}%
, so that let us focus on the case of NMG. To do this, one goes back to the
Fefferman-Graham expansion (\ref{FG}) and uses the leading behavior of the
NMG\ equations of motion. More precisely, it is sufficient to consider the $%
rr$ component of the equations (\ref{eom}) up to the fourth order in the $%
1/r $ expansion. One finally finds%
\begin{equation*}
\gamma ^{ij}T_{ij}^{(\text{ren})}=\frac{l}{16\pi G}\left( 1+\frac{1}{%
2m^{2}l^{2}}\right) \gamma ^{ij}R_{ij}=\frac{c_{L}+c_{R}}{48\pi }R,
\end{equation*}%
that is%
\begin{equation}
c\equiv \frac{c_{L}+c_{R}}{2}=\frac{3l}{2G}\left( 1+\frac{1}{2m^{2}l^{2}}%
\right) .  \label{c}
\end{equation}

This result agrees with the result for the central charge obtained by
different methods: In \cite{Sun} the central charge was obtained as the
central extension of the algebra of asymptotic charges; see \cite{Soda,
Kraus} for the calculation in a quite general three-dimensional theory. In 
\cite{HohmTonni}, on the other hand, the central charge $c$ was obtained by
looking at the Schwarzian derivative term in the anomalous transformation of 
$T_{ij}^{(\text{reg})}$ under $r$-dependent diffeomorphisms. The fact these
computations match supports the interpretation of the tensor (\ref{Tijren})
as the stress-tensor of a CFT$_{2}$ in the boundary of the space.

\subsection{Conserved charges}

The expression for the boundary stress-tensor (\ref{Tijren}) can now be used
to compute conserved charges associated to asymptotic isometries. One is
mainly concerned with the conserved charges associated to asymptotic Killing
vectors $\partial _{t}$\ and $\partial _{\varphi }$, which correspond to the
mass and the angular momentum, respectively. To define the charges it is
convenient to make use of the ADM formalism adapted to the boundary. Then,
the charges are defined by \cite{BY}%
\begin{equation}
Q[\xi ]=\int ds\ \xi ^{i}u^{j}T_{ij}^{(\text{ren})},  \label{carga}
\end{equation}%
where $ds$ is the volume element (i.e. the line element) of the constant-$t$
surfaces at the boundary, $u$ is a unit vector orthogonal to the constant-$t$
surfaces, and $\xi $ is the asymptotic Killing vector.

Here we concentrate in (\ref{generalized}). Applied to such spaces,
calculation (\ref{carga}) yields vanishing conserved charges both for the
mass $Q[\partial _{t}]$ and the angular momentum $Q[\partial _{\varphi }]$.
For instance, considering (\ref{ansatz}), in the large $r$ limit each of the
pieces (\ref{TA})-(\ref{TC}) contributes with a term proportional to $\beta
t/l$. When bringing all the pieces together, one finds%
\begin{equation*}
T_{\varphi \varphi }^{(\text{ren})}=lT_{\varphi t}^{(\text{ren}%
)}=l^{2}T_{tt}^{(\text{ren})}=\frac{c_{L}}{12\pi }h_{+}^{(0)}=0
\end{equation*}%
and the tensor vanishes, as it results proportional to $c_{L}$. Besides,
although each individual contribution to $T_{ij}^{(\text{ren})}$ is
generically non-zero, the contractions $\xi ^{i}u^{j}T_{ij}^{(\text{ren})}$
of each of the three pieces vanishes independently. This yields vanishing
charges%
\begin{equation*}
L_{0}+\overline{L}_{0}\sim Q[\partial _{t}]=0,\quad L_{0}-\overline{L}%
_{0}\sim Q[\partial _{\varphi }]=0.
\end{equation*}%
This generalized the observation made in \cite{Detournay} about the fact
that $h_{++}(t,r,\varphi )$ does not appear in the boundary stress-tensor (%
\ref{TA})-(\ref{TC}), cf. \cite{GiribetLeston}.

\section{Conclusions}

The question remains as to whether well-behaved non-locally AdS$_{3}$
solutions to chiral gravity with finite (non-vanishing) conserved charges
exist. The non-Einstein solutions at the chiral point that obey
Brown-Henneaux boundary conditions found so far happen not to contribute in
a substantial way to the partition function, as the conserved charges
associated to them are zero. One can imagine a plausible scenario in which
non-Einstein spaces do not actually count: For example, even in the case one
finally finds such a non-locally AdS$_{3}$ solution with non-vanishing
charges, one may still ask whether solutions of such type but free of closed
timelike curves and free of naked singularities in the global space do
exist. If non-Einstein spaces free of pathologies do not exist, and if the
Euclidean path integral formulation summing only over smooth geometries
results to be the adequate way of defining the quantum theory, then the
partition function would still be calculable by following the approaches
considered in the literature.

Still far from being able to answer these questions completely, in this
paper we have made two remarks on non-locally AdS$_{3}$ solutions that
asymptote AdS$_{3}$ in massive gravity. These remarks should be taken as a
cautionary note when thinking of extending the chiral gravity theory of \cite%
{CG} by adding the NMG terms to it. First, we have shown that non-Einstein
solutions found in \cite{Detournay} persist when the higher-curvature terms
are added to the gravity action. Such solutions exhibit constant
curvature-invariants, so that it is not totally surprising that they result
resilient under higher-curvature deformation of the action. Furthermore, we
have shown that the addition of the higher-curvature terms also entails the
emergence of new non-Einstein spaces that have no counterpart in TMG, and
which asymptote AdS$_{3}$ for $m^{2}l^{2}$ sufficiently large. In
particular, the solutions we studied include time-independent, axially
symmetric deformations of the extremal BTZ geometry that are not locally
equivalent to it. This implies that the Birkhoff-like theorem proven for TMG
is circumvented in the full massive gravity theory.

\begin{equation*}
\end{equation*}

This work was supported by UBA, CONICET and ANPCyT. The authors thank Alan
Garbarz and Guillem P\'{e}rez-Nadal for discussions and collaboration.

\end{document}